\documentclass[12pt]{llncs}
\usepackage{fullpage}

\begin{document}

\title{Empirical entropy in context}
\author{Travis Gagie}
\institute{Department of Computer Science\\
    University of Toronto\\
    \email{travis@cs.toronto.edu}}
\maketitle

In statistics as in life, many things become clearer when we
consider context.  Statisticians' use of context itself becomes
clearer, in fact, when we consider the past century.  It was
anathema to them prior to 1906, when Markov~\cite{Mar06} proved the weak law of
large numbers applies to chains of dependent events over finite
domains (i.e., finite-state Markov processes).\footnote{Kolmogorov~\cite{Kol36}
later extended this result to Markov processes over infinite
domains.} He published several papers on the statistics of dependent
events and in 1913 gave an example of dependence in language: he
analyzed the first 20\,000 characters of Pushkin's {\it Eugene
Onegin} and found the likelihood of a vowel was strongly affected by
the presence of vowels in the four preceding positions. Many other
examples have been found since, in physics, chemistry, biology,
economics, sociology, psychology --- every branch of the natural and
social sciences. While Markov was developing the idea of Markov
processes, another probability theorist, Borel, was starting an
investigation into examples beyond their scope.  Borel~\cite{Bor09} defined a
number to be normal in base $b$ if, in its infinite $b$-ary
representation, every $k$-tuple occurs with relative frequency \(1 /
b^k\); he called a number absolutely normal if normal in every base.
Using the Borel-Cantelli Lemma, he showed nearly all numbers are
absolutely normal, although his proof was completely
non-constructive. Sierpinski~\cite{Sie17} gave the first example of an absolutely
normal number but his construction is still not known to be
computable.  The first concrete example of a normal number (although
not absolutely normal) was found by Champernowne~\cite{Cha33}, who proved
\(0\,.\,1\,2\,3\,4\,5\,6\,7\,8\,9\,10\,11\,12\dots\) --- the
concatenation of the integers --- is normal in decimal. Champernowne
published his number while still an undergraduate at Cambridge,
where he would certainly have known of Markov processes from
studying economics with Keynes.\footnote{Although later famous as an
economist, Keynes wrote his 1909 dissertation, {\it A
Treatise on Probability}, on statistics; published in 1921, the update references
included eleven works by Markov.}  Champernowne was a close friend
of Turing, who would also have known of Markov processes from
studying statistics with Hardy,\footnote{Like Keynes, Turing wrote
his dissertation on statistics, {\it On the Gaussian Error
Function}.} and Turing~\cite{Tur??} tried to find a concrete example of an
absolutely normal number.\footnote{The question of whether there
exist computable absolutely normal numbers was finally settled
affirmatively in 2002, by Becher and Figueira~\cite{BF02}.} It seems quite
possible the definition of a Markov process influenced Turing's
definition of a universal machine (i.e., Turing
machine),\footnote{Link~\cite{Lin06} discusses this point in his article on the
transmission of Markov's ideas to the West.} and his search for an
absolutely normal number influenced his study of incomputable
numbers.\footnote{Hodges~\cite{Hod06} discusses this point in his review of Copeland's {\it The Essential Turing}.}  After all, Markov processes were good models for the behaviourist psychology popular in the 1930s, Turing is famous for studying computational models of the human mind, and the existence of computable normal numbers shows Turing machines' tapes increase their computational power.

Shannon~\cite{Sha48} made extensive use of Markov processes in his seminal 1948
paper on information theory.  He proposed that any function \(H
(P)\) measuring our uncertainty about a random variable $X$ that
takes on values according to \(P = p_1, \ldots, p_\sigma\) should
have three properties: $H$ should be continuous in the $p_i$; if all
the $p_i$ are equal, \(p_i = 1 / \sigma\), then $H$ should be a
monotonic increasing function of $\sigma$; if a choice should be
broken down into two successive choices, the original $H$ should be
the weighted sum of the individual values of $H$.  He proved the
only function with these properties is his entropy function, \(H (P)
= \sum_{i = 1}^\sigma p_i \log (1 / p_i)\).\footnote{The choice of
the logarithm's base determines the unit; by convention $\log$ means
$\log_2$ and the units are bits --- each being our uncertainty about
a fair coin flip, or a binary digit chosen uniformly at random.}
This axiomatization only elucidated his main results, the Noiseless
and Noisy Coding Theorems, as he did not use it in their proofs. The
Noiseless Coding Theorem, in its simplest form, says the minimum
expected length of a prefix-free code for the value of $X$ is in the
semi-closed interval \([H (P), H (P) + 1)\); notice it cannot be
applied directly when the probability distribution is unknown, as it
is for natural written languages. For this reason, Shannon defined
the entropy of a stationary ergodic Markov process to be,
essentially, the limit as $n$ goes to infinity of \(1 / n\) times
the entropy of the distribution induced by the process over strings
of length $n$; thus, the Noiseless Coding Theorem means that if we
draw a string $s$ of length $n$ from a stationary ergodic Markov
process with entropy $h$, then \(1 / n\) times the expected minimum
length of a prefix-free code for $s$ approaches $h$ as $n$ goes to
infinity.  He fitted zeroth-, first- and second-order Markov
processes to English, gave samples of their output and wrote ``the
resemblance to ordinary English text increases quite noticeably at
each of the above steps'', and ``a sufficiently complex stochastic
process will give a satisfactory representation of a discrete
source'', including ``natural written languages such as English,
German, Chinese.''  It seems Shannon was unaware of both Champernowne's number and another number
--- the concatenation of the primes --- that Copeland and Erd\"{o}s~\cite{CE46}
had proven normal in decimal in 1946.  The existence of such numbers
invalidates the strongest interpretation of Shannon's claim: a
program generating Champernowne's number (e.g., {\tt print "0."; for
i $\geq$ 1 \{print i\}}), for example, cannot be represented as a
(finite-state) Markov process.

Chomsky~\cite{Cho57} argued Markov processes are also inadequate models for
natural language.  For example, he famously claimed that a
probabilistic model cannot determine whether a novel sentence is
grammatical (e.g., ``Colorless green ideas sleep furiously.'' versus
``Furiously sleep ideas green colorless.'') and that Markov
processes, in particular, cannot recognize agreement between
widely-separated words (as in, e.g., ``The man who\dots is here.''
versus ``The man who\dots are here.'', where the ellipses replace
an arbitrarily long verb phrase).\footnote{Chomsky's conclusions
were sweeping: ``the notion `grammatical in English' cannot be
identified in any way with the notion `high order of statistical
approximation to English'\,'' and ``probabilistic models give no
particular insight into some of the basic problems of syntactic
structure''.} His solution, proposed in a series of articles and a
book between 1956 and 1959, was a hierarchy of grammar and language
types
--- regular, context-free, context-sensitive and unrestricted --- in
which, he proved, the set of languages at each level is a proper
superset of the set of languages in the classes below.
This proof should have settled the debate over whether natural
languages can be viewed as coming from Markov processes; it
did in linguistics and psychology, following Chomsky's devastating
review~\cite{Cho59} of Skinner's {\it Verbal Behavior}.  In engineering, however,
despite initial enthusiasm for Chomsky's ideas --- he presented them
at a symposium\footnote{The second day of the symposium included
presentations by Newell and Simon, Chomsky, and Miller, who later
called it ``the moment of conception of cognitive science''\cite{Mil03}.} at MIT
in 1956 organized by the Special Interest Group in Information
Theory, and first published them in the {\it IRE Transactions on
Information Theory} --- the simplifying assumption that sources are
Markovian remains common to this day; in a survey of the first fifty
years of information theory, Verd\'{u}~\cite{Ver98} relegates Chomsky's
opposition to a footnote.

It may be that engineers prefer Markov processes to Chomsky's
grammars because a number of basic problems about grammars are
intractable or incomputable.  The most famous of these is to find
the smallest unrestricted grammar that generates precisely a given
string; since unrestricted grammars are Turing-equivalent, this is
the same as finding the string's Kolmogorov
complexity~\cite{LV97}.\footnote{Assuming \(P \neq \mathit{NP}\), even finding the smallest context-free grammar that generates precisely a given
string is intractable and inapproximable to within a factor of
\(8569 / 8568\)~\cite{CLL+05}.}  Defined independently by Solomonoff~\cite{Sol64} in 1964,
Kolmogorov~\cite{Kol65} in 1965 and Chaitin~\cite{Cha69} in 1969, a string's Kolmogorov
complexity is the length in bits of the shortest program (in a fixed
Turing-equivalent language) that outputs it; according to the
Church-Turing thesis, this is the minimum number of bits needed to
express the string. Notice that, unlike Shannon's entropy,
Kolmogorov complexity is defined for individual strings rather than
sources and requires no probabilistic assumptions. There are two
important facts, however, that limit its usefulness: although
changing the programming language affects the length of the shortest
program by at most an additive constant, that constant may be quite
large (the length in bits of an interpreter for the first
language written in the second language); more importantly, a simple
diagonalization shows Kolmogorov complexity is incomputable and
inapproximable.

In 1976 Lempel and Ziv~\cite{LZ76} proposed an efficiently-computable complexity metric for strings, based on the maximum number of distinct non-overlapping substrings they contain.  As an example, they showed de Bruijn sequences are complex with respect to their metric; a $\sigma$-ary de Bruijn sequence~\cite{deB46} of order $k$ is a $\sigma$-ary string containing every possible $k$-tuple exactly once.  Investigating their complexity metric led Lempel and Ziv to develop their well-known LZ77~\cite{ZL77} and LZ78~\cite{ZL78} compression algorithms.  They used its properties to prove, for example, that LZ78's compression ratio is always asymptotically bounded from above by that of any compression algorithm implementable as a finite-state transducer, regardless of the source.  Nevertheless, Cover and Thomas~\cite{CT91} in their 1991 textbook presented an analysis of LZ78 that assumes the source to be stationary and ergodic, despite noting in an earlier section that ``It is not immediately obvious whether English is a stationary ergodic process.  Probably not!''.  Kosaraju and Manzini~\cite{KM99} introduced another complexity metric, empirical entropy, to re-analyze LZ77 and LZ78 in 1999.  The $k$th-order empirical entropy of a string is its minimum self-information with respect to a $k$th-order Markov source, divided by its length; the self-information of an event with probability $p$ is \(\log (1 / p)\).  They considered families of strings in which the minimum self-information with respect to a Markov source is sublinear in the length, so the empirical entropy approaches 0 as the length increases; Ziv and Lempel's analyses imply LZ77's and LZ78's compression ratios also approach 0, but Kosaraju and Manzini showed the ratios' convergence is asymptotically slower than the empirical entropy's --- so the ratios are not generally within a constant factor of the empirical entropy.\footnote{Specifically, Kosaraju and Manzini proved LZ78's compression ratio is not asymptotically bounded within a constant factor of the $k$th-order empirical entropy for \(k \geq 0\), nor is LZ77's for \(k \geq 1\); the latter is bounded by $8$ times the $0$th-order empirical entropy plus lower-order terms.}

The order of an empirical entropy says how much it depends on the string's ordering.  For example, since a $0$th-order Markov source is just a probability distribution, the $0$th-order empirical entropy is simply the entropy of the normalized distribution of characters, which does not depend on the order of the characters at all.\footnote{Indeed, the earliest bounds we know of in terms of 0th-order empirical entropy, proven by Munro and Spira~\cite{MS76} in 1976, were on the complexity of sorting a multiset.  The best-known algorithm for this problem, splay-sort, is based on the 1985 paper in which Sleator and Tarjan~\cite{ST85} introduced splay-trees and analyzed their performance in terms of 0th-order empirical entropy.}  We can view the $k$th-order empirical entropy \(H_k (s)\) of a string $s$ of length $n$ over an alphabet of size $\sigma$ as our expected uncertainty about a randomly chosen character, given a context of length $k$; it then follows from the Noiseless Coding Theorem that \(H_k (s) n\) is a lower bound on the number of bits needed to encode $s$ with any algorithm that uses contexts of length at most $k$.  Let \(s [i]\) denote the $i$th character of $s$ and consider the following experiment: $i$ is chosen uniformly at random from \(\{1, \ldots, n\}\); if \(i \leq k\), then we are told \(s [i]\); otherwise, we are told \(s [i - k] \cdots s [i - 1]\).  Our expected uncertainty about the random variable \(s [i]\) --- its expected entropy --- is
\[H_k (s) = \left\{ \begin{array}{ll}
\displaystyle \frac{1}{n} \sum_{i = 1}^\sigma n_i \log \frac{n}{n_i}
\hspace{3ex} & \mbox{if \(k = 0\),}\\
& \\ \displaystyle
\frac{1}{n} \sum_{|w| = k} |s_w| \cdot H_0 (s_w)
\hspace{3ex} & \mbox{if \(k \geq 1\).}
\end{array} \right.\]
Here, $n_i$ is the frequency in $s$ of the $i$th character in the alphabet, and $s_w$ is the string obtained by concatenating the characters immediately following occurrences of string $w$ in $s$ --- the length of $s_w$ is the number of occurrences of $w$ in $s$ unless $w$ is a suffix of $s$, in which case it is 1 less.  Notice \(H_{k + 1} (s) \leq H_k (s) \leq \log \sigma\) for all $k$.  For example, if $s$ is the string TORONTO, then
\begin{eqnarray*}
H_0 (s)
& = & \frac{1}{7} \log 7 +
\frac{3}{7} \log \frac{7}{3} +
\frac{1}{7} \log 7 +
\frac{2}{7} \log \frac{7}{2}
\approx 1.84\ ,\\
&& \\
H_1 (s)
& = & \frac{1}{7} \left( \rule{0ex}{2ex}
H_0 (s_\mathrm{N})
+ 2 H_0 (s_\mathrm{O})
+ H_0 (s_\mathrm{R})
+ 2 H_0 (s_\mathrm{T})
\right)\\
& = & \frac{1}{7} \left( \rule{0ex}{2ex}
H_0 (\mathrm{T})
+ 2 H_0 (\mathrm{RN})
+ H_0 (\mathrm{O})
+ 2 H_0 (\mathrm{OO})
\right)\\
& = & 2 / 7 \approx 0.29
\end{eqnarray*}
and all higher-order empirical entropies of $s$ are 0.  This means if someone chooses a character uniformly at random from TORONTO and asks us to guess it, then our uncertainty is about \(1.84\) bits.  If they tell us the preceding character before we guess, then on average our uncertainty is about \(0.29\) bits; if they tell us the preceding two characters, then we are certain of the answer.  Our ability to precisely quantify a string's empirical entropies distinguishes empirical entropy from an earlier notion, stochastic complexity, advocated by Rissanen~\cite{Ris89} in a series of articles and books starting in 1983.  The stochastic complexity (or minimum description length) of a string with respect to a class of sources is the minimum sum of the self-information of the string with respect to a source and the number of bits needed to represent that source.  Stochastic complexity has two theoretical advantages over empirical entropy --- it does not require the sources to be Markovian and it takes into account their complexities --- but is much more complicated and less concrete, since the method of encoding sources is often unspecified.  In any case, although the definition of \(H_k (s)\) does not mention the complexity of the $k$th-order Markov source with respect to which $s$  has minimum self-information, this source can be specified exactly with an \(O (\sigma^{k + 1} \log (n / \sigma^{k + 1}))\)-bit table listing how often each character follows each $k$-tuple in $s$.

Two papers established empirical entropy as a popular complexity metric for strings, at least in the data structures research community.  The first was Manzini's 2001 analysis~\cite{Man01} of the Burrows-Wheeler Transform~\cite{BW94}, in which he proved Burrows and Wheeler's compression algorithm stores $s$ in \(8 H_k (s) n + (\mu + 2 / 25) n + \sigma^k (2 \sigma \log \sigma + 9)\) bits for every \(k \geq 0\) simultaneously, where $\mu$ is a small implementation-dependent constant.\footnote{If arithmetic coding is used in the algorithm's implementation, then \(\mu \approx 1 / 100\).}  Other authors had analyzed the Burrows-Wheeler Transform previously but used ``the hypothesis that the input comes from a finite-order Markov source [which] is not always realistic, and results based on this assumption are only valid on the average and not in the worst case.''  In contrast, Manzini's was a worst-case bound because ``the empirical entropy resembles the entropy defined in the probabilistic setting (for example, when the input comes from a Markov source) [but] is defined for any string and can be used to measure the performance of compression algorithms without any assumption on the input.''\footnote{Kaplan, Landau and Verbin~\cite{KLV??} recently improved Manzini's bound to \(\lambda H_k (s) n + n \log (\zeta (\lambda)) + O (\sigma^{k + 1} \log \sigma)\) bits, where $\lambda$ is any constant greater than 1 and $\zeta$ is the Riemann zeta function.}  The second was Ferragina and Manzini's introduction~\cite{FM05} of compressed full-text indices; exploiting the relationship between the Burrows-Wheeler Transform and suffix array data structures, they showed how to store $s$ in \(5 H_k (s) n + o (n)\) bits such that, given a pattern of length $\ell$, we can find all $\mathrm{occ}$ occurrences of that pattern in $s$ in \(O (\ell + \mathrm{occ} \log^{1 + \epsilon} n)\) time for any \(k \geq 0\) and \(0 < \epsilon < 1\).  This result attracted a lot of attention and, subsequently, there has been a flood of research involving empirical entropy.\footnote{Navarro and M\"{a}kinen~\cite{NM07} recently surveyed dozens of papers on compressed full-text indices.}  Of course, there are still many open questions and probably undiscovered applications; regardless of how much context we have considered, in this case we cannot guess what will come next.

\vfill
\pagebreak

\bibliographystyle{plain}
\bibliography{entropy}

\end{document}